\documentclass[12pt,a4paper,showpacs]{article}
\usepackage{graphicx,multicol}
\usepackage{amsmath, amssymb}
\usepackage{url}
\usepackage{hyperref,rotating}
\begin{document}
\textwidth=135mm
 \textheight=200mm
\begin{center}
{\bfseries Mixed phase effects on high-mass twin stars}
\vskip 5mm
D.E.~Alvarez-Castillo$^{\dag}$\footnote{Email: alvarez@theor.jinr.ru}, 
D.~Blaschke$^{\dag,\ast}$
\vskip 5mm
{\small {\it $^\dag$ Bogoliubov Laboratory of Theoretical Physics, JINR, 141980 Dubna, Russia}\\
{\small {\it $^\ast$ Institute of Theoretical Physics, University of Wroclaw, Poland}}
}
\\
\end{center}
\vskip 5mm
\centerline{\bf Abstract}
Recently it has been found that a certain class of hybrid star equations of state with a large latent heat (strong first order phase transition obtained by a Maxwell construction) between stiff hadronic hadronic and stiff quark matter phases allows for the appearance of a third family of compact stars (including "twins") 
at high mass of $\sim 2~M_\odot$.
We investigate how robust this high-mass twin phenomenon is against a smoothing of the transition which would occur, e.g., due to pasta structures in the mixed phase.
To this end we propose a simple construction of a pasta-like equation of state with a parameter that quantifies the degree of smoothing of the transition and could eventually be related to the surface tension
of the pasta structures.
It is interesting to note that the range of energy densities for the transition as well as the pressure at the 
onset of the transition of this class of hybrid star matter at zero temperature corresponds well to values of the same quantities found in finite temperature lattice QCD simulations for the 1 $\sigma$ region at the 
pseudocritical temperature $T_c=154 \pm 9$ MeV.
The pattern of the speed of sound as a function of energy density is very different.
\vskip 10mm
\section{\label{sec:intro}Introduction}

Neutron stars are dense compact objects where matter exists at extremely high densities. 
For the most massive neutron stars their cores could potentially be composed of quark matter in a deconfined phase where quarks are not longer localized inside nucleons \cite{Alford:2006vz}. 
In this case, as the density increases
towards the center of the star a phase transition between hadronic and quark matter should occur. 
These are the so called hybrid compact stars~\cite{Glendenning} for which a systematic investigation of the relationship between the equation of state (EoS) and the shape of their mass-radius (M-R) diagram has recently been performed by Alford, Han and Prakash (AHP) \cite{Alford:2013aca} for a simplified scheme 
of hybrid EoS. 
The EoS contains all the information of the state of matter at the microscopic level. Phase transitions can happen in an abrupt way, usually described by a Maxwell construction or in a soft, smooth way.
For the description of such smooth transition different interpolation methods have been implemented~\cite{Blaschke:2013rma,Blaschke:2013ana,Alvarez-Castillo:2013spa,Kojo:2014rca}
aiming at capturing its most essential traits.
 On the contrary, a strong first order phase transition appears as a flat plateau in the pressure vs energy density diagram, associated to a latent heat. 
The appeareance of pasta like structures in the transition region has the effect of smoothening this plateau. Here we study these different possibilities.

In this work we have chosen the recently developed models for high mass twins~\cite{Benic:2014jia} 
due to their relevance for the identification of a first order phase transition in the QCD diagram
via astrophysical observations. 
They incorporate realistic elements like excluded volume in the hadronic phase and multiquark interactions 
in the quark matter phase~\cite{Benic:2014iaa}. 
High-mass twins are potentially detectable in future observational missions like the
Neutron Star Interior Composition Explorer (NICER)~\footnote{http://heasarc.gsfc.nasa.gov/docs/nicer/index.html}, 
the Nuclear Spectroscopic Telescope Array (NUSTAR)~\footnote{http://www.nasa.gov/mission-pages/nustar/main} and/or 
the Square Kilometer Array (SKA)~\footnote{http://www.skatelescope.org}.

Bayesian analysis studies are the proper tool to provide estimates of the feasibility of detection~\cite{arXiv:1408.4449,arXiv:1402.0478}. 
Therefore the inclusion of pasta phases is highly important for an effective assessment of identification. 

\section{Hybrid EoS: Maxwell vs. pasta-like}

When the hadron-to quark matter phase transition is accompanied by a large jump in energy density
it entails the formation of rather compact structures, like droplets in a vapour. 
In this situation, surface tension and charge screening become important to make  realistic estimates.  
Most recent microscopical studies based on advanced hadronic and quark matter EoS have confirmed the picture that the Maxwell plateau gets smoothened in a not too dramatic way by pasta structures in the mixed phase, see ~\cite{Yasutake:2014oxa} and references therein. 
Here we suggest a parametrization of the EoS under pasta effects as
\begin{equation}
 \varepsilon( p)=\varepsilon_h( p)f_<( p)+\varepsilon_q( p)f_>( p)~,
\end{equation}
\begin{equation}
 f_{\lessgtr}(p)=\frac{1}{2}\left[1+\tanh\left(\mp~\frac{p-p_c}{\Gamma_s}\right)\right]~,
\end{equation}
where the functions $\varepsilon_h( p)$ and $\varepsilon_q( p)$ are the energy densities in the (h)adronic and (q)uark matter phases, respectively, which result from an inversion of the equations of state 
$p_h(\varepsilon)$ and $p_q(\varepsilon)$. 
The parameter $\Gamma_s$ quantifies the smoothing of the phase transition, i.e. the broadening of the corresponding pressure region from zero (Maxwell transition at $p=p_c$) to a maximum value for which should hold $\Gamma_s \ll p_c$ so that the asymptotic regions, in particular at low pressures, remain unaffected by this construction. 

By varying $\Gamma_s$ we can quantify the robustness of the hybrid star models against the formation of pasta phases.
In particular: How large can this parameter get  without destroying the high-mass twin phenomenon?

\begin{figure}[htb!]
\centering
\includegraphics[width=0.95\textwidth, angle=0]{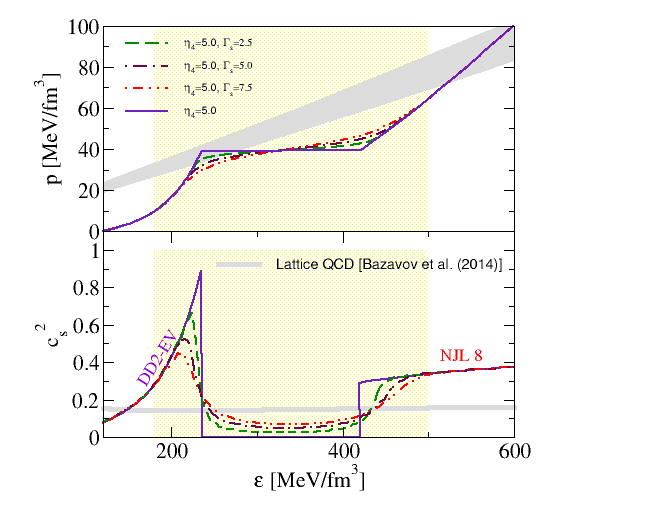}
\caption{Upper panel: Hybrid EoS from Maxwell construction (solid line) between hadronic DD2 with excluded volume and NJL quark matter with higher order interactions. 
Smoothing of the transition increases with the parameter $\Gamma_s$ and is shown by the dashed, 
dash-dotted and dash-double-dotted lines.
Lower panel:  Squared speed of sound for the above EoS. 
The hatched yellow transition region $180 < \epsilon[$MeV/fm$^3] < 500$ has been obtained by lattice studies of hot QCD~\cite{Bazavov:2014pvz}. 
\label{eos_sos}}
\end{figure}

Out of curiosity, we show in Fig.~\ref{eos_sos} the above EoS together with the result of recent lattice QCD studies~\cite{Bazavov:2014pvz} . 
The latter show similarities in the values of pressure and width of the transition region in energy densities, but also striking differences in the behaviour of the speed of sound.
While even a smoothened pasta-like transition is a first order transition at zero (low) temperature, the nature of the deconfinement transition in hot QCD is a crossover.

\section{High-mass twins in the M-R diagram}

\begin{figure}[htb!]
\centering
\includegraphics[width=0.95\textwidth, angle=0]{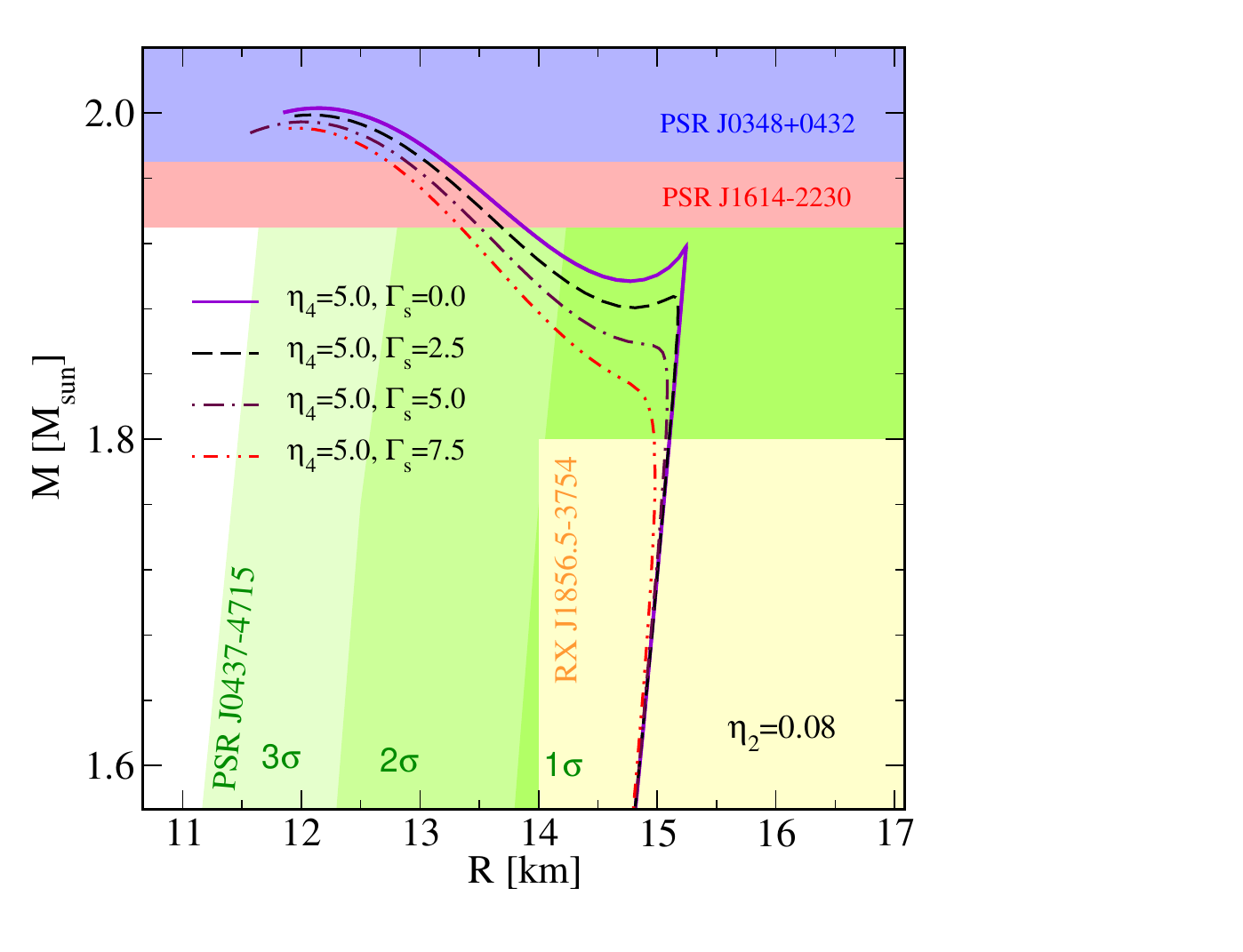}
\caption{Mass-radius relation for hybrid stars with the inclusion of pasta phases at the transition. 
The light blue and light red horizontal regions
correspond to the most massive compact stars detected~\cite{Antoniadis2013,Demorest:2010bx}. 
Radius constraints are shown from Ref.~\cite{Bogdanov2013} by the  light green 1-,2- and 3- $\sigma$
confidence regions and from Ref.~\cite{Hambaryan204} by the light yellow square located in the right 
lower corner.  These support the stiff hadronic EoS required for the high-mass twin star scenario.
\label{MvsR}}
\end{figure}

To compute the mass and radius relation of neutron stars we solve for a set of equations in the framework of General Relativity. 
For a static, non-rotating star, the Einstein equations for spherical symmetry apply and result in the
the Tolman-Oppenheimer-Volkoff equations \cite{Tolman:1939jz,Oppenheimer:1939ne,Shapiro:1983}
\begin{equation}
 \frac{dp( r)}{dr}= - G \frac{(\varepsilon( r)+p( r))(m( r) + 4\pi r^{3}p( r))}{r^{2}(1-2Gm( r)/r)}
,~~~~~~~~~~~  \frac{dm( r)}{dr}=4\pi r^2 \varepsilon( r) ,
\end{equation}
$G$ the gravitational constant and we use units where $\hbar=c=1$. 
This system requires an EoS of the form $p(\varepsilon)$ in order to be solved for the profiles of mass 
$m( r)$, pressure $p( r)$ and energy density $\varepsilon( r)$ as a function of the distance $r$ from the 
center of the star. 
To find a solution it is necessary to choose a central density $\varepsilon( r=0)=\varepsilon_{c}$ 
and to take into account the boundary conditions that $m(r=0)=0$, $m(r=R)=M$ and $p(r=R)=0$ 
which define the radius $R$ and the enclosed gravitational mass $M$ of the star. 
By varying the value of $\varepsilon_{c}$ one obtains $M(\varepsilon_{c})$ and $R(\varepsilon_{c})$,
the parametric form of the mass-radius relation (sequence) which is characteristic for a given EoS. 
In Fig.~\ref{MvsR} we show the results for our class of EoS keeping the parameters of the EoS for hadronic and quark matter phases fixed and varying the smoothing parameter from $\Gamma_s=0$ (Maxwell construction) to $\Gamma_s=7.5$ MeV/fm$^3$, where the there is no separated third family of a stable  hybrid star sequence any more since it has joined the second family of stable neutron stars at a critical value of  $\Gamma_s=5.0$ MeV/fm$^3$. 
Note that for this critical value of the parameter the pressure domain for the phase transition gets
extended to about $\pm 10$ MeV/fm$^3$ around the critical pressure $p_c=40$ MeV/fm$^3$ of the Maxwell transition. 
While for the Maxwell transition case no mixed phase can be realized between quark matter
core and hadronic shell in the hybrid star, we demonstrate here that the smoothing of the transition which leads to the realization of an extended mixed phase will not immediately destroy the high-mass twin phenomenon as a possible observable signature for a strong first order phase transition.

\section{Conclusions}

We have performed a phenomenological study of the robustness of the high-mass twin phenomenon for compact  star $M-R$ relations against the effects of an extended mixed phase due to pasta structures in between the hadronic shell and the quark core of hybrid stars.
We have employed a construction scheme for the pasta-like mixed phase EoS, controlled by 
a parameter which stands for the smoothing of the phase transition region. 
We could show that for not too low values of this parameter twin stars persist. 
Curiosity has driven us to compare the pasta-like $T=0$ EoS for compact stars with the one from lattice QCD for vanishing net baryon density. 
We found that the pressure at the onset of the transition and the transition region of energy densities in hybrid stars both compare well with the corresponding quantities found by lattice QCD in the complementary region of the QCD phase diagram.
However, the softening of the EoS due to the transition, which is measured by a drop in the the speed of sound, is bound to a rather narrow transition region for cold star matter whereas it is washed out over a larger region of energy densities for the crossover transition in hot QCD.
Between these extremes may lie a line of critical endpoints of the spinodal transition in the three-dimensional QCD phase diagram.  

\section{Acknowledgements}
We thank Stefan Typel and Sanjin Benic for providing the necessary hadronic and quark matter EoS 
inputs for this work. 
This research was supported by the Polish NCN under grant No. UMO-2011/02/A/ST2/00306 and by the
Bogoliubov-Infeld Program for collaboration between JINR Dubna and Polish Universities and Institutes.
D.E.A.-C. received support from the STSM programme of the COST Action MP1304 "NewCompStar" under grant No. COST-STSM-MP1304-21192.

\end{document}